\documentclass[aps,pre,showpacs,groupedaddress]{revtex4}
\usepackage{amsfonts}
\usepackage{mathrsfs}
\usepackage{graphicx}
\usepackage{amsmath}
\usepackage{amssymb}
\usepackage{amsbsy}

\begin{document}

\title{Adaptive Thouless-Anderson-Palmer approach to inverse Ising problems with quenched random fields}

\author{Haiping Huang and Yoshiyuki Kabashima}
\affiliation{Department of Computational Intelligence and Systems
Science, Tokyo Institute of Technology, Yokohama 226-8502, Japan}
\date{\today}

\begin{abstract}
The adaptive Thouless-Anderson-Palmer equation is derived for
inverse Ising problems in the presence of quenched random fields. We
test the proposed scheme on Sherrington-Kirkpatrick, Hopfield, and
random orthogonal models and find that the adaptive
Thouless-Anderson-Palmer approach allows surprisingly accurate
inference of quenched random fields whose distribution can be either
Gaussian or bimodal, compared with other existing mean-field
methods.

\end{abstract}

\pacs{02.50.Tt, 02.30.Zz, 75.10.Nr}
 \maketitle

\section{Introduction}
The inverse Ising problem has been intensively studied in
statistical physics and computational biology in the past few
years~\cite{Nature-06,Weigt-2009,Zecchina-10,CM-12}. Such studies
are of huge practical and theoretical relevance. On one hand, the
advent of techniques for multi-electrode recording and microarray
measurement produces high-throughput biological
data~\cite{Stev-2011}. Unveiling the biological mechanism underlying
these experimental data poses a challenging computational problem.
In the inverse Ising problem, one tries to construct a statistical
mechanics description of the original system directly from the data,
and it provides a promising tool for dimensional reduction in
modeling vast amounts of biological data~\cite{Mora-11}. On the
other hand, for guaranteeing the reliability of the obtained
description, it is also necessary to examine the reconstruction
performance of the inverse algorithms numerically and/or
analytically by utilizing artificial data that are generated from a
variety of known Ising spin
models~\cite{Tanaka-1998,Roudi-2009,CM-12,Marsili-2011,Cocco-2011}.

In general, the experimental data are described by $M$ independent
samples
$\{\boldsymbol{\sigma}^{1},\boldsymbol{\sigma}^{2},\ldots,\boldsymbol{\sigma}^{M}\}$
in which $\boldsymbol{\sigma}$ is an $N$-dimensional vector with
binary components ($\sigma_{i}=\pm1$) and $N$ is the system size.
The Ising model provides the least structured model to match the
statistics of the experimental data as
\begin{equation}\label{Ising}
    P_{{\rm
Ising}}(\boldsymbol\sigma)=\frac{1}{Z(\mathbf{h},\mathbf{J})}\exp\left[\sum_{(ij)}J_{ij}\sigma_{i}\sigma_{j}+\sum_{i}h_{i}\sigma_{i}\right],
\end{equation}
where $(ij)$ denotes all distinct spin pairs and the partition
function $Z(\mathbf{h},\mathbf{J})$ depends on $N$-dimensional
fields and $\frac{N(N-1)}{2}$-dimensional couplings. These fields
and couplings are chosen to yield the same first and second moments
(magnetizations and pairwise correlations, respectively) as those
obtained from the experimental data. The inverse temperature
$\beta\equiv1/T$ is absorbed into the strength of fields and
couplings.

Based on magnetizations and correlations, the inference of fields
and couplings of the Ising model is a computationally hard problem
especially for large systems. However, one can resort to mean-field
methods, such as naive mean-field (nMF)~\cite{Kappen-1998},
Thouless-Anderson-Palmer (TAP) equation~\cite{Tanaka-1998},
Sessak-Monasson (SM) expansion~\cite{SM-09}, and Bethe approximation
(BA)~\cite{Mezard-09,Ricci-2012,Berg-2012}, to get an approximate
solution to the inverse problem with computationally feasible costs.
Previous investigations have mostly focused on the inference of the
coupling vector, whereas the inference error of fields has been less
analyzed. In fact, external fields represent intrinsically preferred
directions of $\{\sigma_{i}\}$, which are also very important for
understanding information processing in real neuronal
networks~\cite{Nature-06,Huang-2012pre} and gene interaction
networks~\cite{Zecchina-10} and for predicting protein structures
from sequence data~\cite{Weigt-2009,Zecchina-2011}. Therefore, an
accurate estimation of external fields is also highly desirable.

To this end, we propose the adaptive Thouless-Anderson-Palmer
(adaTAP) approach for the inverse Ising problem and establish the
framework on the basis of Gibbs free energy and Gaussian
approximation. Surprisingly, adaTAP yields a very accurate
estimation of external fields, although some other mean-field
methods are more competitive in predicting couplings. We confirm the
efficiency of adaTAP on three kinds of mean-field models: the
Sherrington-Kirkpatrick (SK) model~\cite{SK-1975}, the
Hopfield~\cite{Amit-1987} model and the random orthogonal model
(ROM)~\cite{Parisi-1995}; other existing mean-field inverse
algorithms are also compared.

The outline of this paper is as follows. The adaptive TAP approach
to the inverse Ising problem with quenched random fields is derived
in Sec.~\ref{sec_adaTAP}. In Sec.~\ref{sec_sim}, extensive numerical
simulations are carried out to test the inference performance of
adaTAP on Hopfield model, SK model and ROM. The comparison with
other existing mean-field methods is also made and discussed.
Concluding remarks are given in Sec.~\ref{sec_Sum}.

\section{Adaptive TAP approach}
\label{sec_adaTAP}
For the Ising model defined in Eq.~(\ref{Ising}), we write the
magnetization-dependent free energy (also termed Gibbs free energy)
as
\begin{equation}\label{gibbs}
G(\mathbf{m})=-\mathbf{h}^{T}\mathbf{m}+{\rm
Extr}_{\boldsymbol{\theta}}\left\{\boldsymbol{\theta}^{T}\mathbf{m}-\ln\sum_{\boldsymbol{\sigma}}e^{\frac{1}{2}\boldsymbol{\sigma}^{T}\mathbf{J}\boldsymbol{\sigma}
+\boldsymbol{\theta}^{T}\boldsymbol{\sigma}}\right\},
\end{equation}
where the Lagrange multiplier vector $\boldsymbol{\theta}$ is
introduced to fix magnetizations at all sites to their thermal
expectation values, i.e., $m_{i}=\left<\sigma_{i}\right>$.
$\mathbf{h}^{T}$ denotes the transpose of a vector $\mathbf{h}$. The
notation ${\rm Extr}$ stands for the extremum with respect to the
corresponding parameters ($\boldsymbol{\theta}$ here). The exact
evaluation of the partition function in Eq.~(\ref{gibbs}) is
computationally difficult for a large system. However, one can
resort to mean-field approximations. We adopt the following
strategy. First, each coupling is multiplied by a real number
$l\in[0,1]$, and the Gibbs free energy can then be expressed by
\begin{subequations}\label{gibbs02}
\begin{align}
\begin{split}\label{gibbs02a}
G(\mathbf{m})&=G(\mathbf{m},l=1)=\int_{0}^{1}dl\frac{\partial
G(\mathbf{m},l)}{\partial l}+G(\mathbf{m},l=0)\\
&\simeq G_{{\rm g}}(\mathbf{m},l=1)-G_{{\rm
g}}(\mathbf{m},l=0)+G(\mathbf{m},l=0),
\end{split}\\
G(\mathbf{m},l)&=-\mathbf{h}^{T}\mathbf{m}+{\rm
Extr}_{\boldsymbol{\theta}}\left\{\boldsymbol{\theta}^{T}\mathbf{m}-\ln\sum_{\boldsymbol{\sigma}}e^{\frac{l}{2}\boldsymbol{\sigma}^{T}\mathbf{J}\boldsymbol{\sigma}
+\boldsymbol{\theta}^{T}\boldsymbol{\sigma}}\right\},\\
\begin{split}
G_{{\rm g}}(\mathbf{m},l)&=-\mathbf{h}^{T}\mathbf{m}+{\rm
Extr}_{\boldsymbol{\theta},\boldsymbol{\Lambda}}\Biggl\{\boldsymbol{\theta}^{T}\mathbf{m}-\frac{1}{2}{\rm
tr}(\boldsymbol{\Lambda}\mathbf{\tilde{C}})\\
&-\ln\int
d\boldsymbol{\sigma}e^{-\frac{1}{2}\boldsymbol{\sigma}^{T}(\boldsymbol{\Lambda}-l\mathbf{J})\boldsymbol{\sigma}
+\boldsymbol{\theta}^{T}\boldsymbol{\sigma}}\Biggr\},\label{gibbs02c}
\end{split}
\end{align}
\end{subequations}
where $\tilde{C}_{ij}\equiv\left<\sigma_{i}\sigma_{j}\right>$, and
we have used Gaussian statistics for the binary spins with
expectation constraints, i.e.,
$\left<\sigma_{i}\sigma_{j}\right>_{{\rm
g}}=\left<\sigma_{i}\sigma_{j}\right>_{{\rm Ising}}$ which are
enforced by a symmetric matrix $\boldsymbol{\Lambda}$. Here, ${\rm
tr}(\mathbf{A})$ denotes the trace of a matrix $\mathbf{A}$. For
simplicity, we assume $\boldsymbol{\Lambda}$ is a diagonal matrix
$\boldsymbol{\Lambda}={\rm diag}(\Lambda_{1},\ldots,\Lambda_{N})$
whose diagonal terms are determined via the extremization of the
corresponding Gibbs free energy. The Gaussian approximation makes
the computation of the partition function tractable. This scheme is
also called the expectation consistence
approximation~\cite{Opper-2005} and was applied to derive the
message-passing algorithm for the perceptron learning
problem~\cite{Kaba-2008}. Conventional Plefka
expansion~\cite{Plefka-1982} truncates the power series expansion of
$G(\mathbf{m},l)$ to second order in $l$, but Eq.~(\ref{gibbs02a})
contains terms of all orders. Note that the third term in
Eq.~(\ref{gibbs02a}) is the Gibbs free energy of non-interacting
Ising spins at fixed magnetizations and can be easily evaluated. The
final expression for the Gibbs free energy reads,
\begin{equation}\label{gibbs03}
\begin{split}
G(\mathbf{m})&\simeq-\frac{1}{2}\mathbf{m}^{T}\mathbf{J}\mathbf{m}-\mathbf{h}^{T}\mathbf{m}+\sum_{i}\mathcal
{H}(m_{i}) +\frac{1}{2}\ln{\rm
det}(\boldsymbol{\Lambda}-\mathbf{J})\\
&-\frac{1}{2}\sum_{i}(1-m_{i}^{2})\Lambda_{i}+\frac{1}{2}\left(N+\sum_{i}\ln(1-m_{i}^{2})\right),
\end{split}
\end{equation}
where $\mathcal
{H}(m_{i})\equiv\frac{1+m_{i}}{2}\ln\frac{1+m_{i}}{2}+\frac{1-m_{i}}{2}\ln\frac{1-m_{i}}{2}$
and $\boldsymbol{\Lambda}$ follows the extremization condition of
Eq.~(\ref{gibbs02c}) with $l=1$,
\begin{equation}\label{lambda}
(\boldsymbol{\Lambda}-\mathbf{J})^{-1}_{ii}=1-m_{i}^{2}.
\end{equation}
Equilibrium values of magnetizations are determined by
$\mathbf{m}_{{\rm eq}}={\rm argmin}_{\mathbf{m}}G(\mathbf{m})$ and
the free energy $F=\min_{\mathbf{m}}G(\mathbf{m})$. A quick
calculation gives the self-consistent equation for $\mathbf{m}$,
\begin{equation}\label{meq}
m_{i}=\tanh\left[h_{i}+\sum_{j}J_{ij}m_{j}-m_{i}\left(\Lambda_{i}-\frac{1}{1-m_{i}^{2}}\right)\right],
\end{equation}
which is exactly the adaptive TAP equation first introduced in
Refs.~\cite{Opper-prl2001,Opper-2001} for the Ising model.
Eq.~(\ref{meq}) can also be derived under other mean-field
approximations~\cite{Raymond-2012,Ricci-2013,Yasuda-2013}. The third
term inside the square bracket of Eq.~(\ref{meq}) forms the Onsager
correction term which requires no prior knowledge of the coupling
statistics, playing an important role in inferring external fields.
$\Lambda_{i}$ in Eq.~(\ref{meq}) is a function of $\{m_{i}\}$
determined by Eq.~(\ref{lambda}). The fixed point of the
self-consistent equation gives $\mathbf{m}_{\rm eq}$. We remark here
that Eq.~(\ref{meq}) can be reduced to the normal TAP equation
obtained from a high-temperature expansion of the Gibbs free
energy~\cite{Plefka-1982,Yedidia-1991,Kaba-2012}, i.e., the third
term inside the square bracket of Eq.~(\ref{meq}) becomes
$-m_{i}\sum_{j}(1-m_{j}^{2})J_{ij}^{2}$ ({\em Onsager reaction
term}) through high-temperature expansion.

To obtain the inference equations for couplings, we use the identity
$\mathbf{H}\mathbf{C}=\mathbf{I}$~\cite{Bray-1979,Tanaka-1998,Yasuda-2009}
where $\mathbf{H}$ is the Hessian matrix of the Gibbs free energy
$H_{ij}\equiv\frac{\partial^{2} G}{\partial m_{i}\partial m_{j}}$
and $\mathbf{C}$ is the connected correlation matrix whose entries
are $C_{ij}\equiv\left<\sigma_{i}\sigma_{j}\right>-m_{i}m_{j}$.
$\mathbf{I}$ is an identity matrix. Magnetizations and correlations
are already given by the experimental data in the inverse Ising
problem. Finally, the inference equation reads
\begin{equation}\label{coup}
J_{ij}=-(\mathbf{C}^{-1})_{ij}+m_{i}(\mathbf{B}^{-1})_{ij}
\end{equation}
for $i\neq j$, where
$B_{ij}\equiv\frac{1}{2m_{i}}\left[\chi_{ij}\right]^{2}$ which
expresses how large the change of $m_{i}$ is given a small
perturbation to $\Lambda_{j}$ and we define
$\boldsymbol{\chi}=(\boldsymbol{\Lambda}-\mathbf{J})^{-1}$. The
expression for $B_{ij}$ is derived by using the Sherman-Morrison
formula~\cite{Recipe-2007}. A small perturbation $\Delta\Lambda_{j}$
to $\Lambda_{j}$ will lead to a corresponding change of $m_{i}$
according to Eq.~(\ref{lambda}), which is described by the following
equation:
\begin{equation}\label{Bmat}
[\boldsymbol{\chi}^{-1}+\Delta\boldsymbol{\Lambda}_{j}]_{ii}^{-1}=\chi_{ii}-\frac{\Delta\Lambda_{j}\chi_{ij}^{2}}{1+\Delta\Lambda_{j}\chi_{ij}}\\
=1-(m_{i}+\Delta m_{i})^{2},
\end{equation}
where the Sherman-Morrison formula is used in the first equality and
the notation $\Delta\boldsymbol{\Lambda}_{j}$ means that only $j$-th
diagonal term of matrix $\Delta\boldsymbol{\Lambda}$ is non-zero and
equal to $\Delta\Lambda_{j}$. Noting that both $\Delta m_{i}$ and
$\Delta\Lambda_{j}$ are small, one can obtain
$B_{ij}\equiv\frac{\partial
m_{i}}{\partial\Lambda_{j}}=\frac{1}{2m_{i}}\left[\chi_{ij}\right]^{2}$
by using Eq.~(\ref{lambda}) once again. After couplings are
reconstructed, external fields are inferred as
\begin{equation}\label{field}
h_{i}=\tanh^{-1}(m_{i})-\sum_{j}J_{ij}m_{j}+m_{i}\left(\Lambda_{i}-\frac{1}{1-m_{i}^{2}}\right).
\end{equation}
To predict the coupling, we need to solve the adaTAP equation
Eq.~(\ref{lambda}). An iterative scheme is proposed as follows.
\begin{description}
  \item[Step 1.] Let $t=0$ and initialize $J_{ij}=-(\mathbf{C^{-1}})_{ij}$,
  $\Lambda_{i}=(1-m_{i}^{2})^{-1}$ for all $(ij)$ and $i$ respectively.
  \item[Step 2.] At $t$, set $t'=0$, $\tilde{\boldsymbol{\Lambda}}^{t'=0}=\boldsymbol{\Lambda}^{t}$,
  $\boldsymbol{\chi}=(\boldsymbol{\Lambda}^{t}-\mathbf{J}^{t})^{-1}$.
  \begin{description}
    \item[Step 2.1.] $t'\leftarrow t'+1$, update
    $\tilde{\Lambda}^{t'}_{i}=\tilde{\Lambda}^{t'-1}_{i}+\Delta\tilde{\Lambda}_{i}$
    where
    $\Delta\tilde{\Lambda}_{i}=\frac{1}{1-m_{i}^{2}}-\frac{1}{\chi_{ii}}$
    for all $i$. After update of each $\tilde{\Lambda}^{t'}_{i}$,
    $\boldsymbol{\chi}$ needs to be updated simultaneously as
    $\chi_{kl}^{{\rm new}}=\chi_{kl}^{{\rm old}}-\frac{\chi_{ki}^{{\rm old}}\Delta\tilde{\Lambda}_{i}\chi_{il}^{{\rm old}}}{1+\Delta\tilde{\Lambda}_{i}\chi_{ii}^{{\rm old}}}$
    derived by using the Sherman-Morrison formula.
    \item[Step 2.2] Until
    $|\Delta\tilde{\Lambda}_{i}|<\epsilon_{\Lambda}$ for all $i$, then assign
    $\boldsymbol{\Lambda}^{t}=\tilde{\boldsymbol{\Lambda}}^{t'}$, $\boldsymbol{\chi}^{t}=\boldsymbol{\chi}$ and go
    to \textbf{Step 3}. Otherwise, if $t'<t'_{{\rm max}}$, go to \textbf{Step 2.1}, else return UN-CONVERGED.
  \end{description}
  \item[Step 3.] $t\leftarrow t+1$, update
  $J_{ij}^{t}=-(\mathbf{C}^{-1})_{ij}+m_{i}(\mathbf{B}^{-1})_{ij}$ where
$B_{ij}\equiv\frac{1}{2m_{i}}\left[\chi^{t-1}_{ij}\right]^{2}$,
until $|J_{ij}^{t}-J_{ij}^{t-1}|<\epsilon_{J}$ for all $(ij)$, then
go to \textbf{Step 4}. Otherwise, if $t<t_{{\rm max}}$, go to
\textbf{Step 2}, else return UN-CONVERGED.
  \item[Step 4.] Infer $h_{i}$ according to Eq.~(\ref{field}) for
  all $i$.
\end{description}
In step 2.1, the step size $\Delta\tilde{\Lambda}_{i}$ for updating
$\tilde{\Lambda}_{i}$ can be derived by using Eq.~(\ref{lambda}) and
the Shermon-Morrison formula, which gives
$\chi_{ii}-\frac{\Delta\tilde{\Lambda}_{i}\chi_{ii}^{2}}{1+\Delta\tilde{\Lambda}_{i}\chi_{ii}}
=1-m_{i}^{2}$.  In the iterative scheme, we set the parameters
$t_{{\rm max}}=t'_{{\rm max}}=1000$, and
$\epsilon_{\Lambda}=\epsilon_{J}=10^{-4}$. In practice, we find that
both $\boldsymbol{\Lambda}$ and $\mathbf{J}$ in our simulations
shown below converge in tens of steps when the temperature is not
very low. The computational complexity of this iterative scheme is
dominated by the inverse of the matrix (e.g., $\mathbf{C}$ or
$\mathbf{B}$), keeping the same order as that of other mean-field
methods.


To compare performances of different mean-field inverse algorithms,
we define the inference error for couplings and fields,
respectively, as
\begin{subequations}\label{error}
\begin{align}
\Delta_{J}&=\left[\frac{2}{N(N-1)}\sum_{i<j}(J_{ij}^{*}-J_{ij}^{{\rm
true}})^{2}\right]^{1/2},\\
\Delta_{h}&=\left[\frac{1}{N}\sum_{i}(h_{i}^{*}-h_{i}^{{\rm
true}})^{2}\right]^{1/2},
\end{align}
\end{subequations}
where $J_{ij}^{*}$ ($h_{i}^{*}$) is the inferred coupling (field)
and $J_{ij}^{{\rm true}}$ ($h_{i}^{{\rm true}}$) is the true one.

\section{Numerical simulations}
\label{sec_sim}

We evaluate the inference performance of the adaTAP approach on
three mean-field models with either Gaussian distributed or bimodal
distributed random fields. For the SK model, each entry of the
coupling matrix are independently drawn at random from a Gaussian
distribution with zero mean and variance $1/N$. In the Hopfield
model, the coupling is constructed according to Hebb's rule, i.e.,
$J_{ij}=\frac{1}{N}\sum_{\mu=1}^{P}\xi_{i}^{\mu}\xi_{j}^{\mu}$ where
$P$ random Gaussian patterns $\{\boldsymbol{\xi}^{\mu}\}$ are stored
in the network. $\xi_{i}^{\mu}$ are independent Gaussian random
variables with zero mean and unit variance. We also test our method
on ROM whose coupling matrix is constructed as $\mathbf{J}=\mathcal
{O}^{T}\mathbf{D}\mathcal {O}$ where $\mathcal {O}$ is an orthogonal
matrix chosen with the Haar measure~\cite{Parisi-1995,Dean-2003}.
$\mathbf{D}={\rm diag}(\lambda_{1},\ldots,\lambda_{N})$ and
$\lambda$ follows a distribution
$\rho(\lambda)=\alpha\delta(\lambda-1)+(1-\alpha)\delta(\lambda+1)$.

\begin{figure}
\centering
          \includegraphics[bb=21 17 294 218,width=7.5cm]{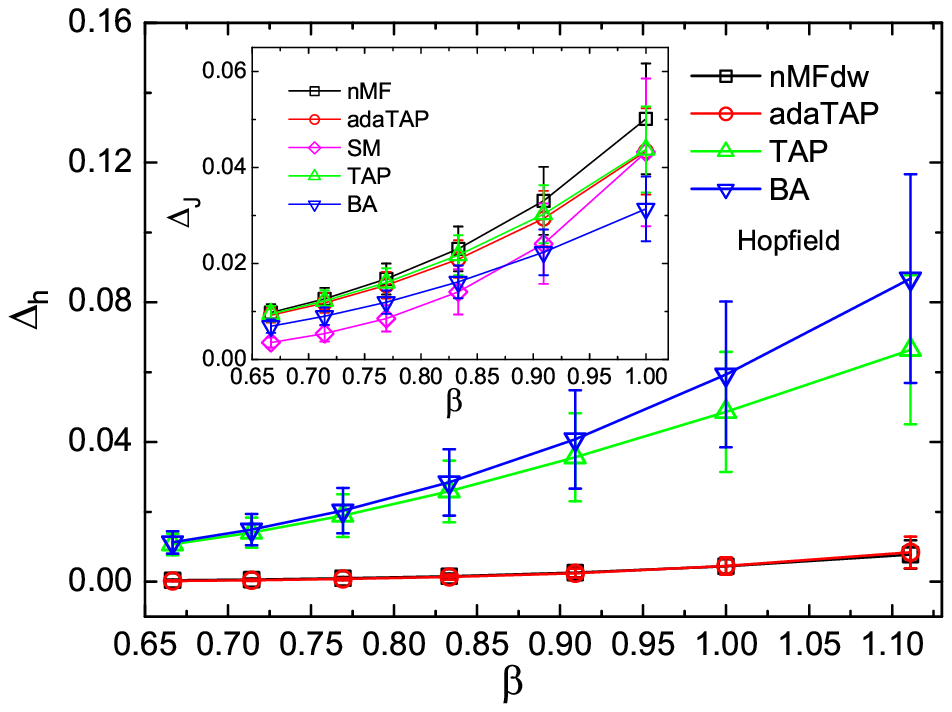}
     \vskip .05cm
     \includegraphics[bb=21 16 293 213,width=7.5cm]{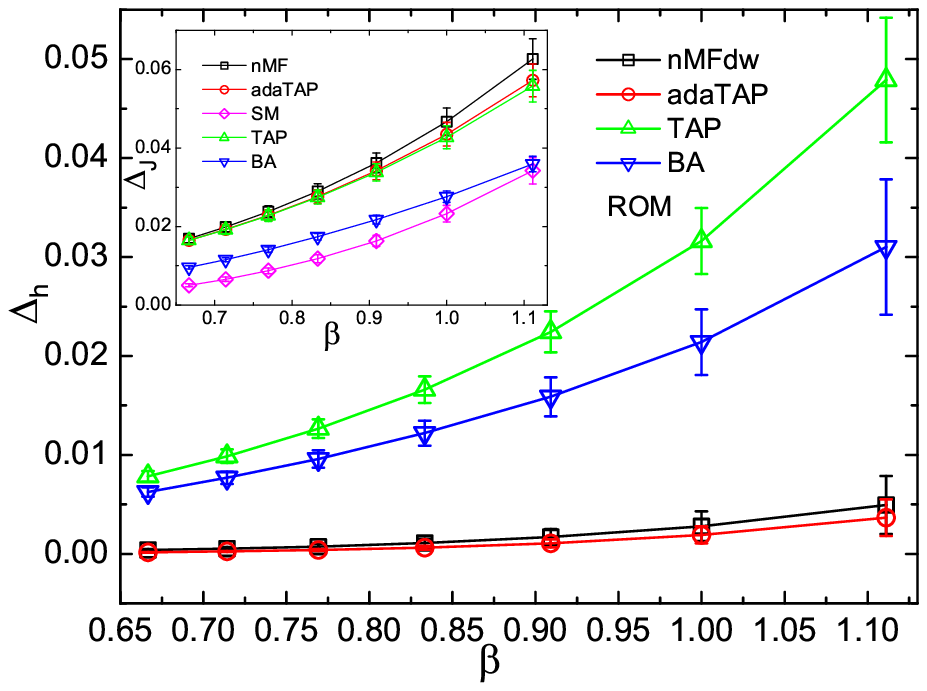}\vskip .05cm
     \includegraphics[bb=22 16 292 216,width=7.5cm]{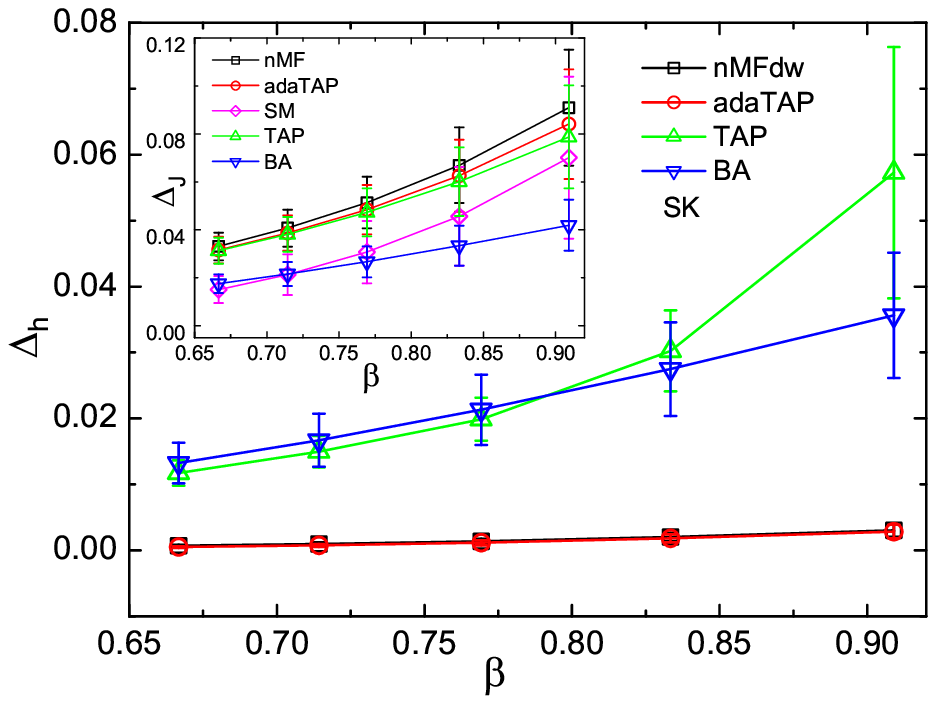}\vskip .05cm
  \caption{(Color online)
     Inference performances of adaTAP on Hopfield, random orthogonal and SK models with Gaussian distributed random fields, compared with those obtained by other existing mean-field methods. Magnetizations and correlations
     used to infer fields and couplings
     are calculated through exact exhaustive enumeration on networks of size $N=15$. Each data marker is the average over $20$ random realizations for which $\sigma_{h}^{2}=0.01$.
      (a) Results for the Hopfield
     model with $P=3$. (b) Results for ROM with $\alpha=0.6$. (c) Results for the SK model.
   }\label{Perf}
 \end{figure}

To collect the data of magnetizations and correlations, we use the
exact enumeration on small-size systems of $N=15$, which produces
noise-free data for predicting the underlying parameters. In this
case, $M=2^{N}$. Inference results of adaTAP on these three tested
models are compared with those obtained by other mean-field methods.
For comparison, we briefly describe the other four existing
mean-field methods for the inverse Ising problem. The couplings
between spin $i$ and $j$ ($i\neq j$) are inferred as follows:
\begin{subequations}\label{mfeq01}
\begin{align}
J_{ij}^{{\rm nMF}}&=-(\mathbf{C^{-1}})_{ij},\\
 J_{ij}^{{\rm
TAP}}&=\frac{2(\mathbf{C^{-1}})_{ij}}{-1-\sqrt{1-8m_{i}m_{j}(\mathbf{C^{-1}})_{ij}}},\\
J_{ij}^{{\rm SM}}&=J_{ij}^{{\rm nMF}}+J_{ij}^{{\rm
ind}}-\frac{C_{ij}}{L_{i}L_{j}-C_{ij}^{2}},\\
   J^{{\rm BA}}_{ij}&=-\tanh^{-1}\Biggl[\frac{1}{2(\mathbf{C}^{-1})_{ij}}(a_{ij}-b_{ij})
   -m_{i}m_{j}\Biggr],
\end{align}
\end{subequations}
where $J_{ij}^{{\rm
ind}}=\frac{1}{4}\ln\left[\frac{(1+\tilde{C}_{ij})^{2}-(m_{i}+m_{j})^{2}}{(1-\tilde{C}_{ij})^{2}-(m_{i}-m_{j})^{2}}\right]$,
$a_{ij}=\sqrt{1+4L_{i}L_{j}(\mathbf{C}^{-1})_{ij}^{2}}$,
$b_{ij}=\sqrt{\left(a_{ij}-2m_{i}m_{j}(\mathbf{C}^{-1})_{ij}\right)^{2}-4(\mathbf{C}^{-1})_{ij}^{2}}$
and $L_{i}=1-m_{i}^{2}$. After couplings are inferred, fields can be
predicted using the following equations:
\begin{subequations}\label{mfeq02}
\begin{align}
h_{i}^{{\rm nMF}}&=\tanh^{-1}(m_{i})-\sum_{j\neq i}J^{{\rm nMF}}_{ij}m_{j},\label{nmfH}\\
 h_{i}^{{\rm TAP}}&=\tanh^{-1}(m_{i})-\sum_{j\neq i}J^{{\rm TAP}}_{ij}m_{j}\nonumber\\
 &+m_{i}\sum_{j\neq i}(J^{{\rm TAP}}_{ij})^{2}(1-m_{j}^{2}),\\
 h_{i}^{{\rm BA}}&=\tanh^{-1}(m_{i})-\sum_{j\neq i}\tanh^{-1}\left(t_{ij}f(m_{j},m_{i},t_{ij})\right),
\end{align}
\end{subequations}
where $t_{ij}=\tanh J^{{\rm BA}}_{ij}$ and
$f(x,y,t)=\frac{1-t^{2}-\sqrt{(1-t^{2})^{2}-4t(x-yt)(y-xt)}}{2t(y-xt)}$. Since SM expansion has large inference errors
for predicting fields even when considering up to the third order in
the small correlation expansion~\cite{SM-09}, we would not show its
field inference performances for the temperature range we consider.

We first examine the inference performance of adaTAP on mean-field
models, where quenched random fields are drawn independently at
random from a Gaussian distribution with zero mean and variance
$\sigma_{h}^{2}$. As displayed in Fig.~\ref{Perf} (a) for the
Hopfield model, adaTAP shows slightly better performance than the
TAP approach in coupling constructions, whereas the SM expansion has
the best performance at high temperatures and the BA has the best
one at low temperatures. Regarding field inference, adaTAP performs
much better than other methods in the entire temperature range under
consideration. However, if we incorporate an effective self-coupling
(diagonal weight)
$J_{ii}=\frac{1}{1-m_{i}^{2}}-(\mathbf{C^{-1}})_{ii}$~\cite{Kappen-1998}
into the inference equation~(\ref{nmfH}), nMF with diagonal weights
(nMFdw) will achieve the nearly same accuracy with adaTAP in
predicting external fields, although adaTAP still gives a bit lower
inference error. This also holds for the other two mean-field
models. Note that nMF without diagonal weights definitely gives a
highest inference error among all mean-field methods compared here.
As the temperature becomes sufficiently low, adaTAP ceases to
converge within $t_{{\rm max}}$ or $t'_{{\rm max}}$, thus becoming
unable to predict couplings and fields. To infer a model with
quenched random fields, TAP and BA will also have no solution at low
enough temperatures. We also performed simulations with a larger $\sigma_{h}^{2}$ (e.g.,
$\sigma_{h}^{2}=0.1$), and it is observed that the field inference
performance deteriorates and adaTAP fails to converge at a higher
temperature for some samples compared to the case with a smaller
field variance. Fig.~\ref{Perf} (b) shows inference results for ROM
with the random orthogonal coupling matrix. Although adaTAP behaves
slightly worse than TAP for inferring couplings, it produces
surprisingly accurate estimates of external fields in the entire
temperature range in Fig.~\ref{Perf} (b). Note that the inference
accuracy obtained by other mean-field methods (except nMFdw) can be
further improved by at least one order of magnitude by using adaTAP
when the random fields are Gaussian distributed. For coupling
inferences of the SK model (see Fig.~\ref{Perf} (c)), the
performance of adaTAP lies between those of nMF and TAP, while the
SM expansion gives a more accurate prediction than other methods at
high temperatures.

\begin{figure}
\centering
          \includegraphics[bb=19 13 285 215,width=7.5cm]{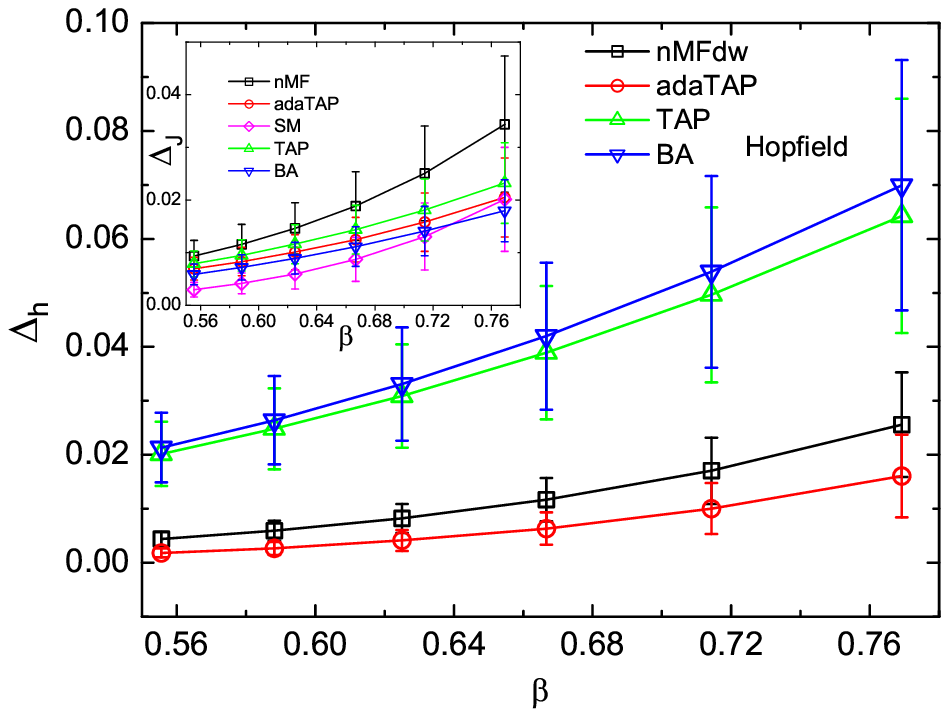}
     \vskip .05cm
     \includegraphics[bb=20 18 281 212,width=7.5cm]{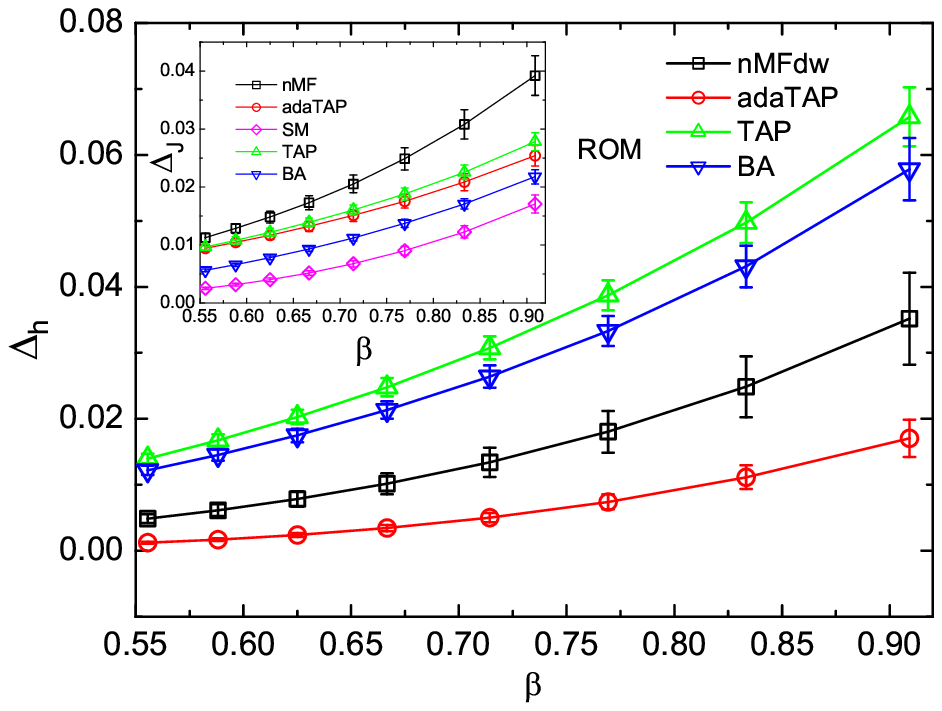}\vskip .05cm
     \includegraphics[bb=19 18 281 211,width=7.5cm]{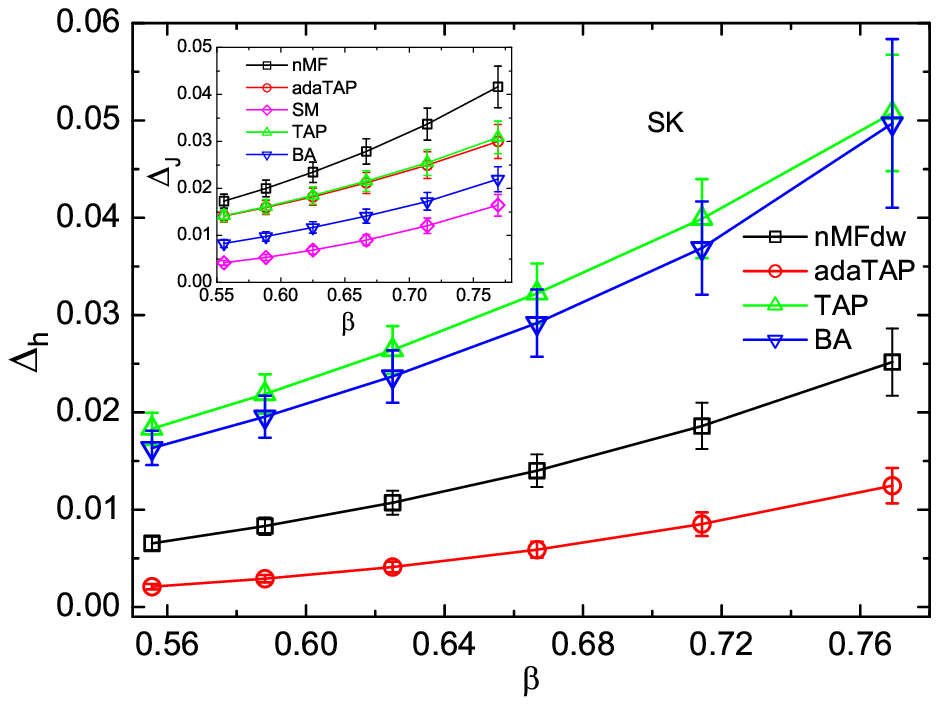}\vskip .05cm
  \caption{(Color online)
     Inference performances of adaTAP on Hopfield, random orthogonal and SK models with bimodal distributed random fields, compared with those obtained by other existing mean-field methods. Magnetizations and correlations
     used to infer fields and couplings
     are calculated through exact exhaustive enumeration on networks of size $N=15$. Each data marker is the average over $20$ random realizations for which $h_{0}=0.3,p=0.6$.
      (a) Results for the Hopfield
     model with $P=3$. (b) Results for ROM with $\alpha=0.6$. (c) Results for the SK model.
   }\label{Perfb}
 \end{figure}

\begin{figure}
\centering
          \includegraphics[bb=22 19 286 220,width=7.5cm]{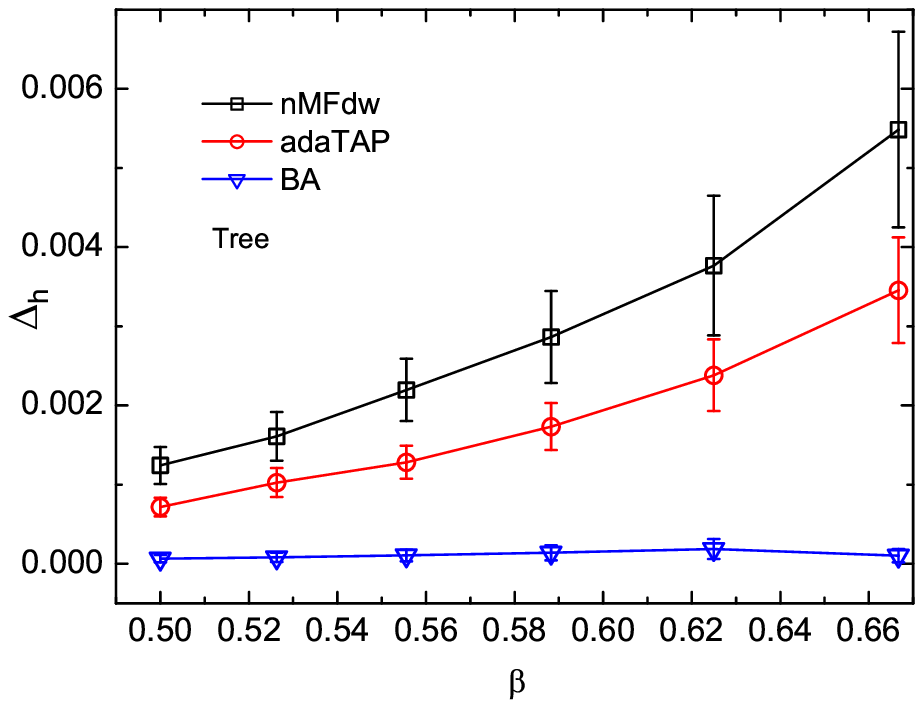}
  \caption{(Color online)
     Inference performances of adaTAP on a tree ($N=22$), compared with nMFdw and BA (exact method on a tree model). Each data marker is the average over $10$ random realizations.
   }\label{tree}
 \end{figure}

Fortunately, the superiority of adaTAP for field inference is also
true when the random field is bimodal distributed, i.e.,
$p_{h}(h)=p\delta(h-h_{0})+(1-p)\delta(h+h_{0})$. Its performance is
shown in Fig.~\ref{Perfb} with $p=0.6, h_{0}=0.3$. The improvement
of the field prediction by adaTAP is evident in this case, even
compared to nMFdw. In adaTAP, we still assume zero diagonal
couplings, however, the third term inside the square bracket of
Eq.~(\ref{meq}) provides an adaptive Onsager correction to the nMF
approximation, playing the same key role with diagonal couplings in
inferring external fields. In this adaptive manner, lower inference
error of fields and couplings is achieved compared to nMFdw.
Interestingly, adaTAP can even perform better than TAP in
predicting couplings for certain ranges of temperatures in this
case.

\begin{figure}
\centering
          \includegraphics[bb=17 17 288 214,width=7.5cm]{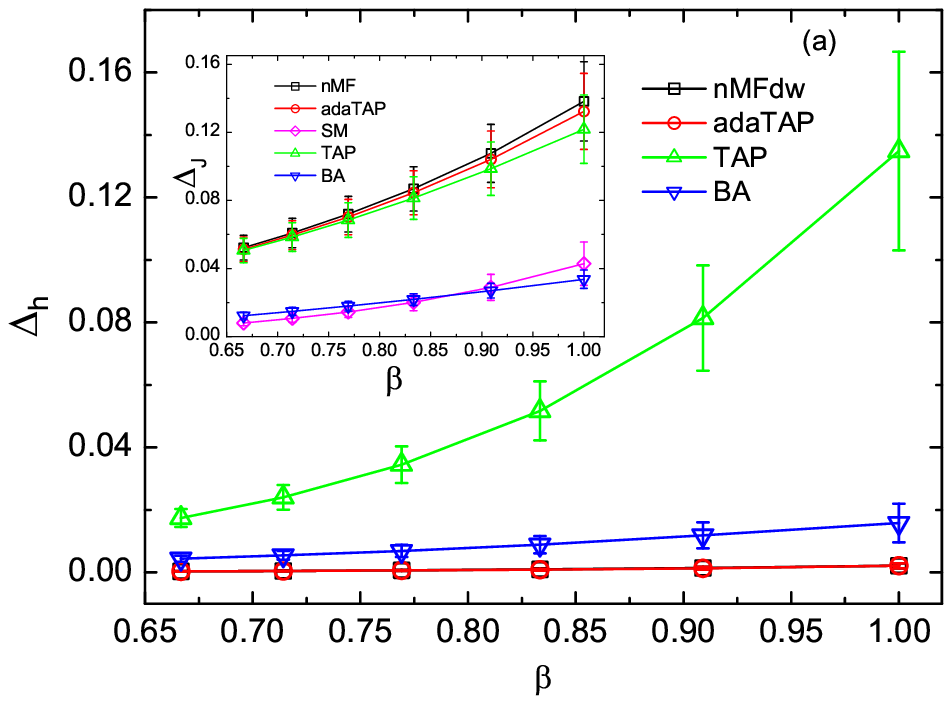}
     \vskip .05cm
     \includegraphics[bb=24 16 287 218,width=7.5cm]{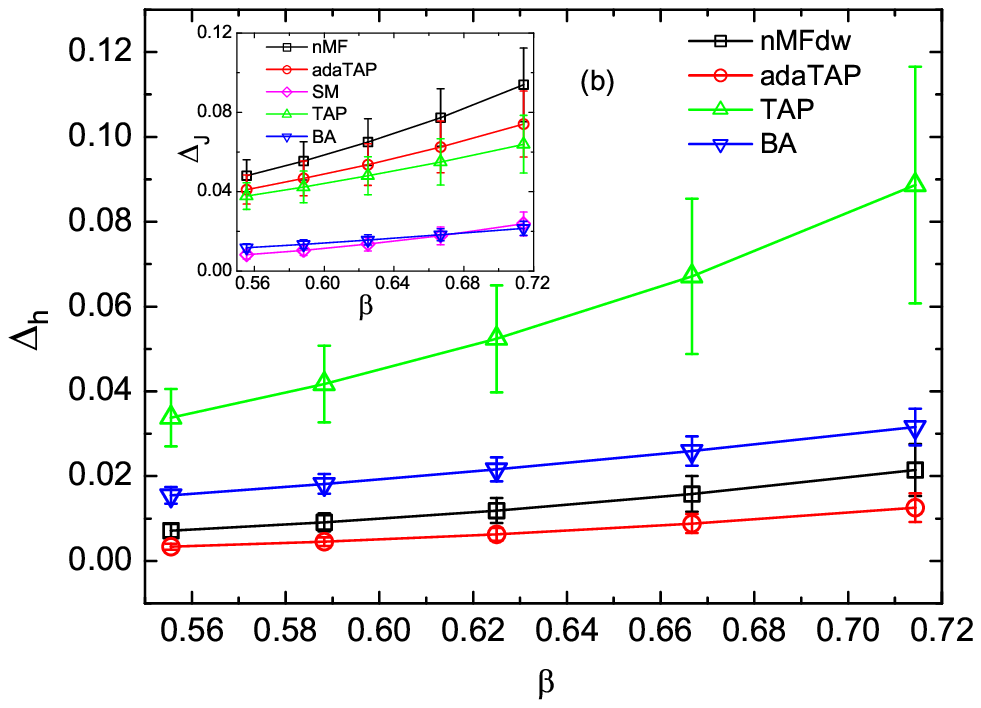}\vskip .05cm
  \caption{(Color online)
     Inference performances of adaTAP on diluted SK model ($p_{d}=0.4$), compared with those obtained by other existing mean-field methods. Magnetizations and correlations
     used to infer fields and couplings
     are calculated through exact exhaustive enumeration on networks of size $N=15$. Each data marker is the average over $20$ random realizations.
      (a) Results for the Gaussian distributed random fields with $\sigma_{h}^{2}=0.01$. (b) Results for the bimodal distributed random fields with $h_{0}=0.3,p=0.6$.
   }\label{Perfc}
 \end{figure}

All the models investigated so far are of the fully connected type. For examining the capability to deal with another extreme of sparsely
connected networks,  the proposed scheme is also tested on a tree model.  Our proposed scheme performs better than other methods
except BA which is exact on a tree and gives very accurate inference
both on the couplings and fields. The result is shown in
Fig.~\ref{tree}. A tree of size $N=22$ is constructed, such that
each node inside the tree has degree equal to $3$, and to mimic an
infinite Bethe lattice, we generate the external fields for the
boundary spins as $\hat{h}_{i}=h_{i}+\sum_{k\in\partial i\backslash
j}h_{k\rightarrow i}$, where $j$ is the only spin inside the tree
connected to the boundary spin $i$ and the cavity field
$h_{k\rightarrow i}$ is randomly chosen from a population dynamics
for an infinite Bethe lattice~\cite{cavity-2001}. Couplings and
fields ($h_{i}$ for the boundary spins) for the tree follow Gaussian
distributions $\mathcal {N}(0,1)$ and $\mathcal {N}(0,0.01)$
respectively. Magnetizations and correlations are calculated by
using susceptibility propagation algorithms~\cite{Huang-2012pre}.  However, in
real applications, for example, a typical neuronal network of size around $N=100$
is not strongly diluted with an exact tree structure, therefore, our
method is expected to still give good estimates of external fields.  To confirm this point, we test adaTAP on a diluted SK model, where each
non-zero Gaussian distributed coupling is present with a predefined
probability $p_{d}$. The Gaussian distribution has zero mean and
variance $1/c$ with $c=p_{d}N$. As shown in Fig.~\ref{Perfc}, the
adaTAP still performs better than other mean-field methods
(including nMFdw) in field inference, which is much more evident when
random fields are bimodal distributed. 

Although adaTAP, TAP, and the BA will have no solution in inferring
mean-field models with quenched random fields at low temperatures,
adaTAP does outperform other existing mean-field methods compared
here to infer quenched random fields if it converges, as shown for a
wide range of temperatures on the Hopfield, random orthogonal and (diluted) SK
models. We conclude that the power of adaTAP for inverse Ising
problems resides in its remarkable accuracy in predicting external
fields, especially for the case where there is a single dominant
state in phase space.
\section{Conclusion}
\label{sec_Sum}

In summary, we propose the adaTAP approach for inverse Ising
problems and show its striking performance for inferring external
fields in mean-field models. As far as the field inference is
concerned, adaTAP is rather satisfactory, compared to other
mean-field methods. Furthermore, an accurate inference of external
fields in the Ising model is able to provide us with insights into
the mechanism underlying high-throughput data either coming from
biological experiments or from large
databases~\cite{Nature-06,Weigt-2009,Zecchina-2011}. The proposed adaTAP
approach for the inverse Ising problem is expected to have applications
in real data analyses (e.g., neural data, or sequences in the
protein databases), in combination with other mean-field methods.



\begin{acknowledgments}
We thank the referees for their helpful comments and suggestions.
This work was partially supported by the JSPS Fellowship for Foreign
Researchers (Grant No. $24\cdot02049$) (HH) and JSPS KAKENHI Nos.
$22300003$ and $22300098$ (YK).
\end{acknowledgments}



\end{document}